\documentclass[showpacs,twocolumn,preprintnumbers,amsmath,amssymb,epsf,superscriptaddress]{revtex4}
\usepackage{txfonts}
\usepackage{graphicx}
\usepackage{lipsum}
\usepackage{nicefrac}
\usepackage{color}
\usepackage{hyperref}
\usepackage{float}
\begin{document}
\title{Three particle Hyper Entanglement: Teleportation and Quantum Key Distribution }
\author{Chithrabhanu P}
\affiliation{Physical Research laboratory, Navarangpura, Ahmedabad, India - 380009}
\author{Aadhi A}
\affiliation{Physical Research laboratory, Navarangpura, Ahmedabad, India - 380009}
\affiliation{IIT Gandhinagar, Chandkheda, Ahmedabad, India-382424}
\author{Salla Gangi Reddy}
\affiliation{Physical Research laboratory, Navarangpura, Ahmedabad, India - 380009}
\author{Shashi Prabhakar}
\affiliation{Physical Research laboratory, Navarangpura, Ahmedabad, India - 380009}
\affiliation{IIT Gandhinagar, Chandkheda, Ahmedabad, India-382424}	
\author{G. K. Samanta}
\affiliation{Physical Research laboratory, Navarangpura, Ahmedabad, India - 380009}
\author{Goutam Paul}
\affiliation{Cryptology \& Security Research Unit, R. C. Bose Centre for Cryptology \& Security, Indian Statistical Institute, Kolkata, India - 700108} 
\author{R.P.Singh}
\affiliation{Physical Research laboratory, Navarangpura, Ahmedabad, India - 380009}
\date{\today}    
\vspace{-10mm} 
\begin{abstract}
We present a scheme to generate three particle hyper-entanglement utilizing polarization and orbital angular momentum (OAM) of a photon. We show that the generated state can be used to teleport a two-qubit state described by the polarization and the OAM. The proposed quantum system has also been used to describe a new efficient quantum key distribution (QKD) protocol. We give a sketch of the experimental arrangement to realize the proposed teleportation and the QKD.
\end{abstract}
\maketitle
\section{Introduction}
  With entanglement between two quantum bits, protocols have been demonstrated for teleporting an unknown quantum state \cite{bennett1993}, super dense coding of information \cite{bennett1992} and secure communication \cite{ekert1991}. An arbitrary qubit can be teleported from one particle to another with the use of  an entangled pair of particles, which had been experimentally verified in different quantum systems \cite{Bouwmeester1997,riebe}. However, distinguishing all the four Bell states of the photonic qubits has remained a fundamental difficulty in achieving 100\% teleportation. In the first demonstration of the teleportation with photons \cite{Bouwmeester1997}, only one of the four Bell states was able to distinguish from the others. Thus, the efficiency of teleportation was limited to 25\%. Later on, a complete Bell state measurement was demonstrated with non-linear interaction of photons\cite{kim}. Even though they could separate all the four Bell states, the efficiency was reduced because of the non-linear process involved. 

In recent years, complete Bell state analysis has been proposed with the use of hyper-entanglement \cite{wei}, where the two photons are entangled in an additional degree of freedom (DOF) along with polarization. This method was utilized to increase the channel capacity of super dense coding \cite{barreiro2008}. This was done by projecting each hyper-entangled photon to four single particle two-qubit Bell states. Nevertheless, hyper-entanglement assisted Bell state analysis is not of much use in teleportation since it requires projecting the unknown state and one of the EPR particle state to any of the four Bell states. On the other hand, a hyper-entangled pair of particles can teleport a higher dimensional quantum state using hyper-entangled-Bell-state analysis which was described using Kerr non-linearity \cite{Sheng1}. There had been a number of studies which use spin-orbit states of light for quantum information processing \cite{Aolita,Souza,Santos,Borges,Lixiang}. Khoury and Milman \cite{Khoury2011} proposed a spin to orbit teleportation scheme with 100\% efficiency that uses the OAM entanglement between the two photons and a spin-orbit Bell state analysis (SOBA).

In this article, we describe a three particle entangled state which finds applications in teleporting two qubits simultaneously and implementing an efficient key distribution protocol. In Section~\ref{sc.2} we give a description of the proposed state. Along with the mathematical form of the state we give a schematic for the state preparation. The experimental procedure for the generation of the proposed state is given in Section~\ref{sc.3}. The state can be utilized to teleport two qubits using two SOBAs and 16 unitary transformations as given in 
Section~\ref{sc.4}. Experimental schemes for realizing the $C_{NOT}$ gates and SOBA have also been given in Section~\ref{exp}. In Section \ref{sc.5} we describe a new QKD protocol using the new state, which is more efficient than the traditional Ekert protocol. Finally, we conclude in Section~\ref{sc.6}. 

\section{Description of the proposed state}\label{sc.2}
We describe a system of particles in such a way that one particle is entangled to all other particles in different degrees of freedom. Let us consider a system consisting of three photons where photon \textbf{2} is entangled with photons \textbf{1} and \textbf{3} in different degrees of freedom namely OAM and polarization respectively. The polarization state of the photon \textbf{1} and the OAM state of the photon \textbf{3} are arbitrary or unknown.

 Since the OAM of a photon is expressed in infinite dimensional Hilbert space, one can have higher dimensional entangled states. We take an arbitrary two dimensional subspace of the infinite dimensional OAM basis as $\{\vert l\rangle, \vert l'\rangle \}$.  
 
The described state can be prepared using a pair of Hadamard and $C_{NOT}$ gates  in different DOFs which correspond to the polarization and the OAM.
\begin{figure}
  \begin{center}
    \includegraphics[width=3in]{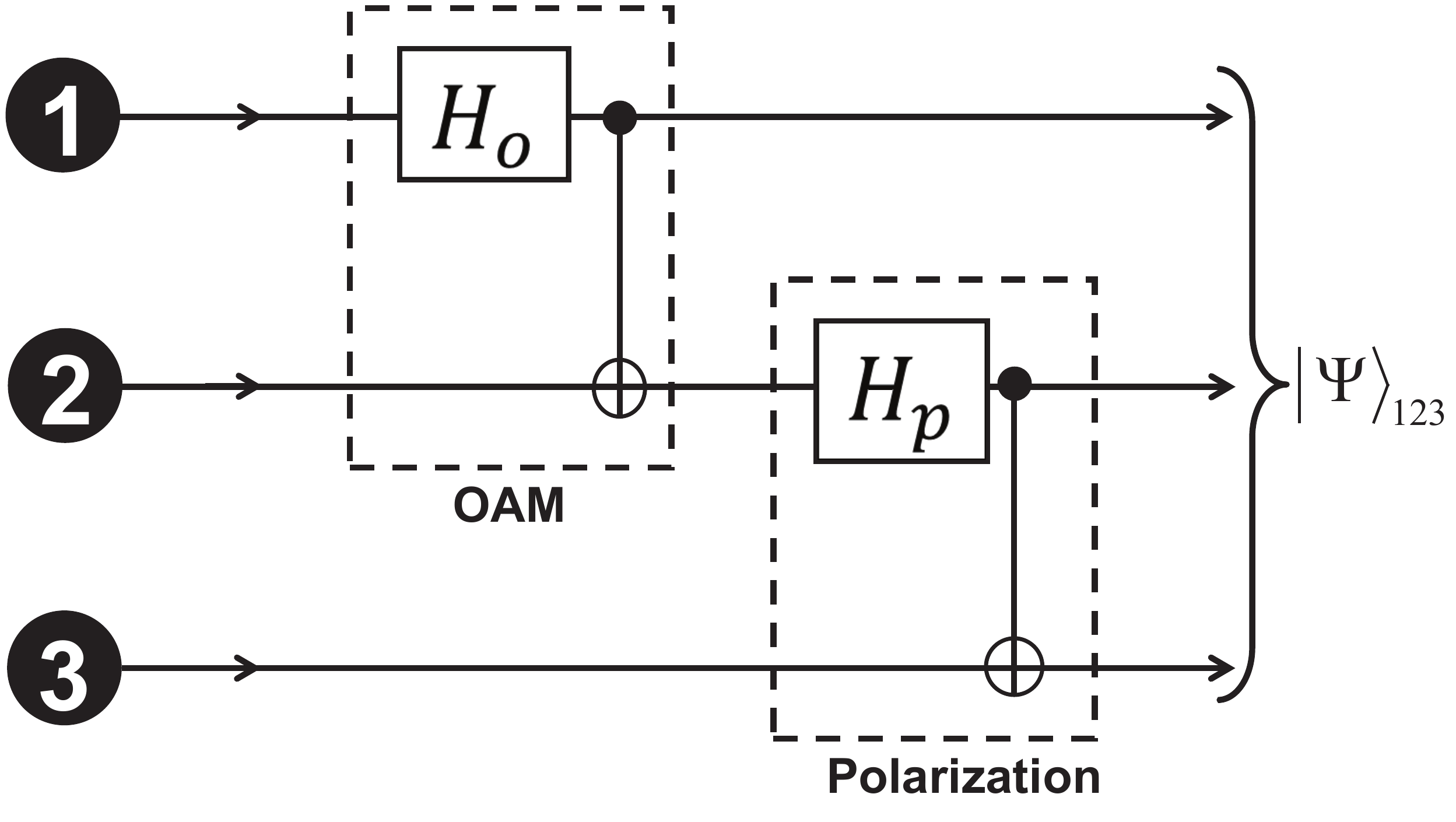}
    \caption{ Schematic diagram for preparation of the initial state.  Box named  ``OAM" contains a Hadamard gate ($H_o$) and a $C_{NOT}$ gate acting on OAM basis and ``polarization" box contains a Hadamard ($H_p$) and $C_{NOT}$ gates acting on polarization basis. \label{fg.2}}
  \end{center}
\end{figure}
The initial states of three particles can be written as 
\begin{eqnarray}
 \vert 1\rangle = \vert \xi_p\rangle_1 \vert l\rangle_1\ ; \
 \vert 2\rangle = \vert H\rangle_2 \vert l\rangle_2\ ; \
 \vert 3\rangle = \vert H\rangle_3 \vert \chi_{o}\rangle_3,
\end{eqnarray}
where $\vert \xi_p\rangle$ is the unknown state of polarization of the photon \textbf{1}, $ \vert\chi_{o}\rangle$ is the unknown  OAM state of the photon \textbf{3}, $ \vert H\rangle $ represents horizontal and $  \vert V\rangle $ represents vertical polarization states of a photon. The schematic diagram for preparation of the initial state is given in Fig.~\ref{fg.2}. To entangle photons \textbf{1} and \textbf{2} in OAM,  a Hadamard gate ($H_o$) on photon \textbf{1} and subsequent $C_{NOT}$ gate on photon \textbf{1} and \textbf{2} are applied. Both gates act in the OAM degree of freedom. Similarly, another Hadamard ($H_p$)  acting on photon \textbf{2} and a $C_{NOT}$ between photons \textbf{2} and \textbf{3} entangle them in polarization. Here, both the gates are acting on the polarization states of photons. 
After performing the gate operations mentioned above, the final state becomes, 

\begin{equation} 
\label{2}
\vert \Psi\rangle_{123} = \vert \xi_p\rangle_1 \otimes(\vert l,l'\rangle_{12}+\vert l',l \rangle_{12}) \otimes (\vert HV\rangle_{23}+\vert VH\rangle_{23} )\vert\chi_{o}\rangle_3.
\end{equation}

\section{Experimental scheme for state preparation}\label{sc.3}
\begin{figure}[h]
  \begin{center}
    \includegraphics[width=3in]{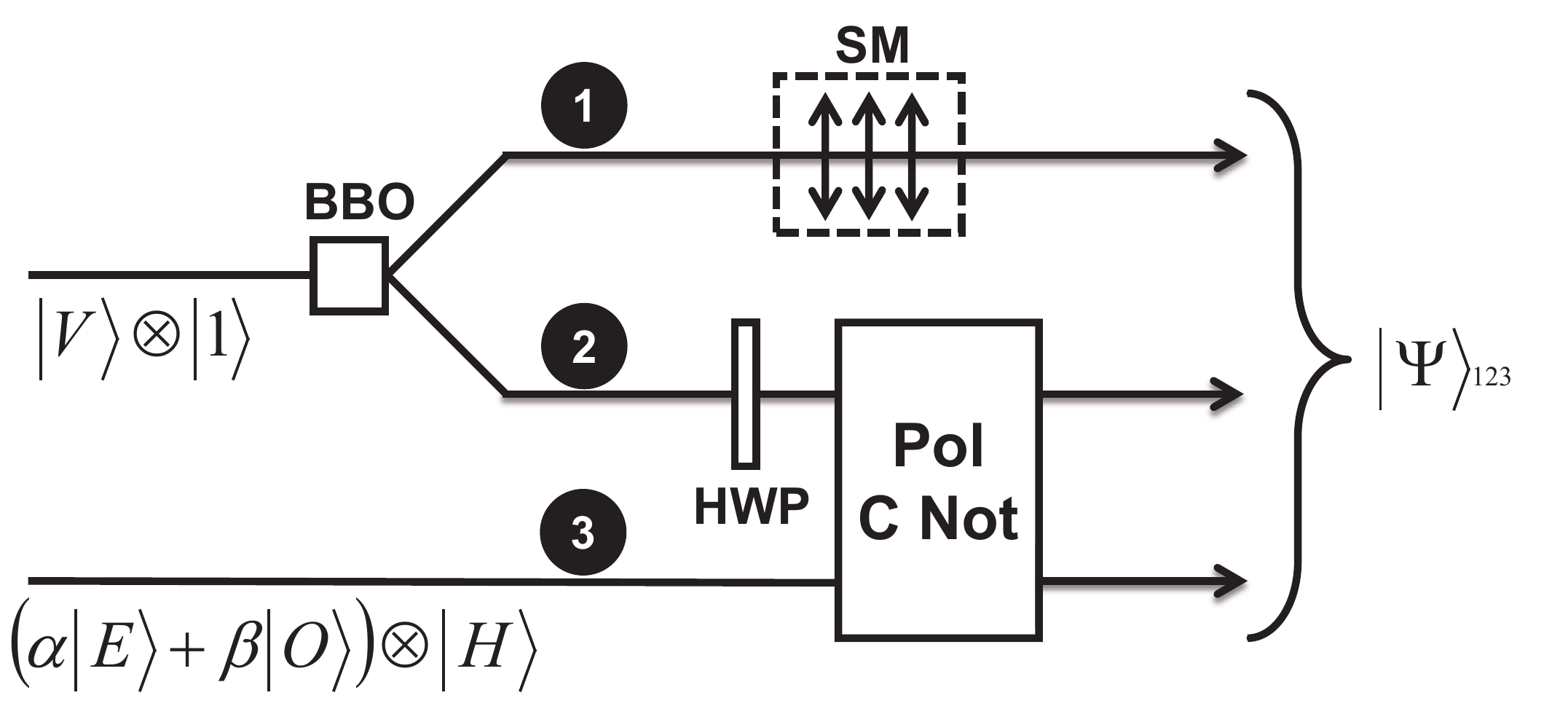}
    \caption{Schematic experimental set up for the state preparation. SM-  Simon-Mukunda gadget, HWP- half wave plate, BBO- second order nonlinear crystal (Beta Barium Borate) . }\label{fg.3}
  \end{center}
\end{figure}

 We are proposing a method in which initially two photons are entangled in the OAM and third photon is in a pure state of OAM and polarization. By polarization gate operations on one of the entangled photons and the independent photon, one can arrive at the described state. 
Experimental scheme for the generation of the state is given in Fig.~\ref{fg.3}.

To generate the described state, we start with a Type I spontaneous parametric down conversion (SPDC) of light in a second order nonlinear crystal that gives a pair of photons entangled in the OAM \cite{Walborn2004}. A vertically polarized optical vortex beam of azimuthal index 1 has been considered as the pump. The state corresponding to the pair of photons produced by the SPDC of this beam is given by
\begin{equation}
\label{1}\vert\Psi\rangle_{12} = \sum_{m=-\infty}^{+\infty} c_m \vert m\rangle_1 \vert 1-m \rangle_2 \otimes \vert H\rangle_1\vert H\rangle_2
\end{equation} 

Note that the experimental realization of the quantum gates in the OAM  basis \{ $\vert 0\rangle, \pm\vert 1\rangle \, \pm\vert 2\rangle ...$ \} are not straightforward. Moreover, in the teleportation, it is easier to project the state of a photon to one of its four spin-orbit Bell state if we use even/odd basis. Thus for the ease of experimental realization, we reduce the infinite dimensional entangled state to a simple two-qubit entangled state by grouping all the even and the odd OAM states and rewrite the expression for the OAM state in Eq.~\eqref{1} as
\begin{eqnarray}
\label{split}
\sum_{m=-\infty}^{+\infty} c_{m} (\vert m\rangle_1 \vert 1-m \rangle_2) \nonumber 
 =   \sum_{k=-\infty}^{+\infty} c_{2k} (\vert 2k\rangle_1 \vert 1-2k \rangle_2) + \\
 \sum_{k=-\infty}^{+\infty} c_{2k+1} (\vert 2k+1\rangle_1 \vert -2k \rangle_2).
\end{eqnarray}

We define a transformation from the general OAM space to the
even-odd OAM space as
$$g(\vert u \rangle_1 \vert v \rangle_2) =
f(\vert u \rangle_1)f(\vert v \rangle_2), \mbox{ where}$$
\begin{equation}
f(\vert x \rangle)
= \left\{ \begin{array}{ll}
\vert E \rangle, & \mbox{ for even $x$};\\
\vert O \rangle, & \mbox{ for odd $x$}.
\end{array} \right.
\end{equation}

Applying $g$ on the left-hand-side of Eq.~\eqref{split} yields
$\frac{1}{\sqrt{2}}(\vert E \rangle_1 \vert O \rangle_2 +
\vert O \rangle_1 \vert E \rangle_2)$.

From the conservation of OAM, we have 
$\sum_{k=-\infty}^{+\infty} (c_{2k})^2 = \sum_{k=-\infty}^{+\infty} (c_{2k+1})^2= \frac{1}{2}\sum_{m=-\infty}^{+\infty} (c_{m})^2 = \frac{1}{2}$.

Thus, we can rewrite Eq.~\eqref{1} as
\begin{equation}
\vert\Psi\rangle_{12} = \frac{1}{\sqrt{2}}\left(\vert E\rangle_1 \vert O \rangle_2 +\vert O\rangle_1 \vert E \rangle_2 \right) \otimes \vert H\rangle_1\vert H\rangle_2.
\end{equation} 

Note that the photons are entangled in the even/odd OAM states. Experimentally, this transformation can be implemented by post selection in the even/odd OAM basis. All the OAM operations or measurements must be performed in even/odd basis using OAM sorter \cite{Leach}. This maps the general OAM state to even/odd basis which is mathematically represented by operator “$g$”. Let the photon \textbf{1} pass through a Simon-Mukunda polarization gadget  which can convert its polarization to any arbitrary state \cite{simon,Reddy} and the photon \textbf{2} pass through a half wave plate at $\frac{\pi}{8}$. Thus polarization state of the photon \textbf{1} is encoded as the unknown state $a\vert H\rangle_1 +b \vert V\rangle_1 $ and the state of the photon \textbf{2} is encoded as $\frac{1}{\sqrt{2}} ( \vert H\rangle_2+ \vert V\rangle_2)$. Action of HWP on the photon \textbf{2} is a Hadamard operation. Thus, the state becomes
\begin{eqnarray}
\vert\Psi\rangle_{12} & = & \frac{1}{\sqrt{2}}\left(\vert E\rangle_1 \vert O \rangle_2 +\vert O\rangle_1 \vert E \rangle_2 \right) \left(a\vert H\rangle_1 +b \vert V\rangle_1\right)\otimes \nonumber \\
& & \frac{1}{\sqrt{2}} ( \vert H\rangle_2+ \vert V\rangle_2).
\end{eqnarray}
Now, consider the photon \textbf{3} with unknown superposition  state of OAM  in the even/odd basis and with definite state polarization. Its state can be expressed as,
\begin{equation}
\vert\Psi\rangle_3 = (\alpha \vert E\rangle_3 + \beta \vert O \rangle_3)\otimes \vert V\rangle_3.
\end{equation}
A polarization $C_{NOT}$ gate is applied on photon \textbf{2} (control) and \textbf{3} (target). This operation leads to a polarization entanglement between photons \textbf{ 2} and \textbf{3}. Thus, the three particle state becomes
\begin{eqnarray}
\label{4} 
\nonumber \vert\Psi\rangle_{123} = \frac{1}{2}\left(a\vert H\rangle_1 +b \vert V\rangle_1\right) \left(\vert E\rangle_1 \vert O \rangle_2 +\vert O\rangle_1 \vert E \rangle_2 \right) \\
( \vert HV\rangle_{23}+ \vert VH\rangle_{23}) (\alpha \vert E\rangle_3 + \beta \vert O \rangle_3). 
\end{eqnarray}
This is in the same form as the proposed three particle 
entangled state described in Section~\textbf{2}, Eq. ~\eqref{2}. 
    
\section{Simultaneous teleportation of two qubits using the new state}\label{sc.4}
\begin{figure*}[!ht]
  \begin{center}
    \includegraphics[width=4in]{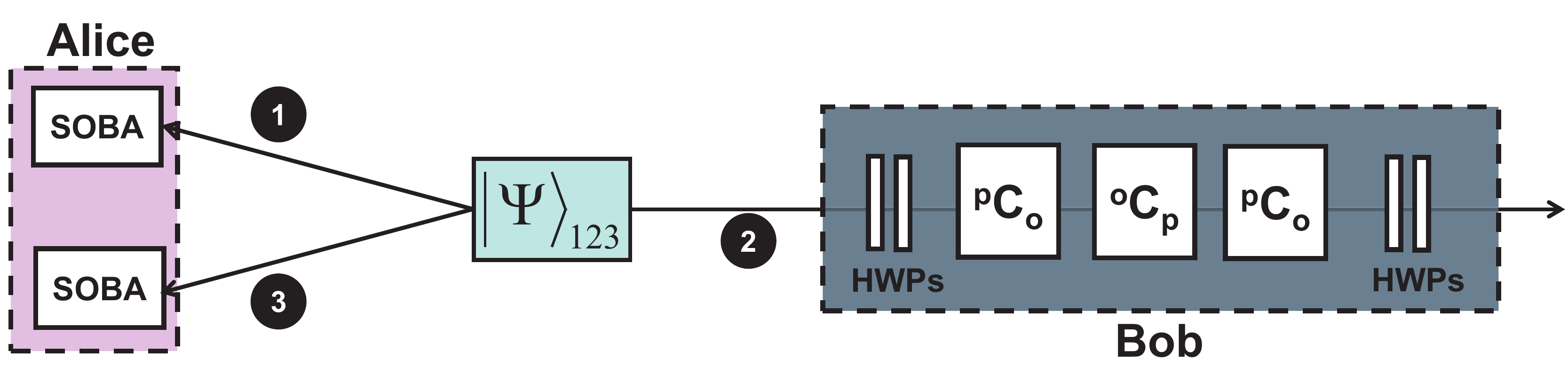}
    \caption{(Colour online) Schematic for the teleportation of spin-orbit qubits shared by photons \textbf{1} and \textbf{3}. HWPs - half wave plates. }\label{fg.4}
  \end{center}
\end{figure*}
Teleportation of $N$ qubits, in general, needs $N$ pairs of entangled photons. On the other hand, teleportation of  $d$-dimensional quantum state needs $d$-dimensional entanglement and one has to do a joint measurement in $d^2$ dimensional Bell basis \cite{pati}. However, teleportation schemes were proposed for a $d$ dimensional OAM state using  Bell filter and quantum scissors \cite{goyal,Goyal2013,babichev2003}.  
 

 With the entangled state presented above, we propose an efficient scheme for teleporting  a two-qubit state distributed in different DOFs. The schematic diagram is given in Fig.~\ref{fg.4}. The polarization state of photon \textbf{1} and the OAM state of photon \textbf{3} form a four dimensional unknown two-qubit state
\begin{equation}
\label{8}\vert \Psi\rangle_u = \left(a\vert H\rangle_1 +b \vert V\rangle_1\right) \otimes (\alpha \vert E\rangle_3 + \beta \vert O \rangle_3),
\end{equation}
which needs to be teleported. We can teleport the combined state to a single particle, photon \textbf{2}.    

The unknown polarization (spin angular momentum) and the entangled OAM state of photon \textbf{1} and entangled polarization and the unknown OAM state of photon \textbf{3} in Eq.~\eqref{4}, can be projected to individual spin-orbit Bell states by SOBA. Alice performs the SOBA on the photons \textbf{1} and \textbf{3} and projects the state of both the particles to the corresponding spin-orbit Bell states. The SOBA, for which the OAM is in the even/odd basis, can be achieved by Mach-Zehnder interferometers involving an OAM sorter and polarizing beam splitters \cite{Khoury2011}. 

The states can be defined as 
\begin{eqnarray}
  \nonumber \psi^{\pm} =&  \frac{1}{\sqrt{2}}\left( \vert H, E\rangle \pm \vert V, O\rangle\right),\\
  \phi^{\pm} =& \frac{1}{\sqrt{2}}\left( \vert H, O\rangle \pm \vert V, E\rangle\right).
\end{eqnarray} 
 The SOBA operation for photon \textbf{1} can be represented by a polarization controlled OAM $C_{NOT}$ ($^pC_o $) gate and a polarization Hadamard ($H_p$) gate operation with subsequent detection. Similarly, the SOBA for photon \textbf{3} is an OAM controlled polarization $C_{NOT}$ ($^oC_p $) gate and OAM Hadamard ($H_o$) gate operation with subsequent detection. Experimental schemes for these single particle two-qubit $C_{NOT}$ gates and SOBA are given in 
Section~\ref{exp}.

By substituting the single particle spin-orbit Bell states into a three particle wave function in Eq.~\eqref{4}, we get
\begin{eqnarray}
\vert\Psi\rangle_{123} = \frac{1}{4} \sum_{x,y=0}^{1}\sum_{x',y'=0}^{1} \vert\Phi^{xy}\rangle_1\vert\Phi^{x'y'}_{q}\rangle_3\otimes 
\Gamma_{x,y,x',y'} \\
  \left(\alpha\vert H\rangle_2 +\beta \vert V\rangle_2\right) (a \vert E\rangle_2 + b \vert O \rangle_2),\nonumber
\end{eqnarray}
where $\vert\Phi^{xy}\rangle_1 $ and $ \ \vert\Phi^{x'y'}_{q}\rangle_3  $ are  single particle spin-orbit Bell states corresponding to the photons \textbf{1} \& \textbf{3} respectively. $ \Gamma_{x,y,x',y'}  $ is a transformation matrix for the spin-orbit state of photon \textbf{2}. It has been introduced for the compact representation of the 16 states of photon \textbf{2} given in Table~\ref{tb.1}. 
After the SOBA measurements on the photons \textbf{1} and \textbf{3}, state of the photon \textbf{2} becomes
\begin{equation}
\vert \Psi(D_{\vert\Phi^{xy}\rangle_1\vert\Phi^{x'y'}_{q}\rangle_3 })\rangle_2 = \frac{1}{4} \Gamma_{x,y,x',y'} 
\left(\alpha\vert H\rangle_2 +\beta \vert V\rangle_2\right)(a \vert E\rangle_2 + b \vert O \rangle_2),
\end{equation}
 the state corresponding to a detection of $ \vert\Phi^{xy}\rangle_1 \vert\Phi^{x'y'}_{q}\rangle_3 $ in SOBAs.

The states corresponding to all possible SOBA outcomes are given in the first and the second columns of the Table~\ref{tb.1}. Note that the information encoded in the polarization state of the photon \textbf{1} is transferred to the OAM state of the photon \textbf{2} and the information encoded in the OAM state of the photon \textbf{3} is transferred to the polarization state of the photon \textbf{2}. So, to get the initial state, Bob has to do a swapping ($U_{SWAP}$) between the OAM and the polarization along with other unitary transformations which use only polarization operations. These operations can be implemented with the standard wave plates which makes the experimental realization of this method more feasible.

\begin{figure}[h]
  \begin{center}
    \includegraphics[width=3in]{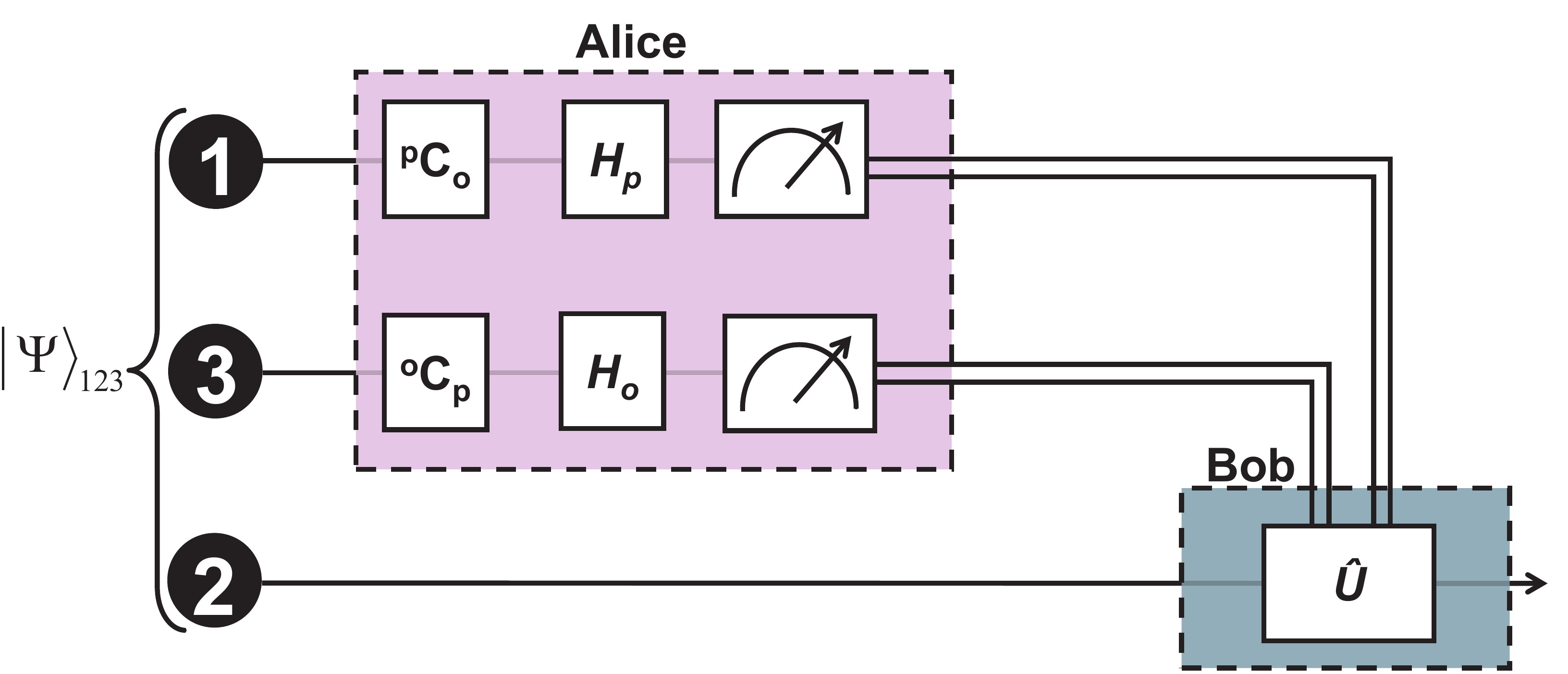}
    \caption{(Color online) Circuit diagram for proposed teleportation scheme. Single/double lines - quantum/classical communication channels, arrow - measurement. $\hat{U}$ - unitary transformation given in Table~\ref{tb.1}}\label{fg.5} 
  \end{center}
\end{figure}

There are 16 possible outcomes for Alice measurements which demand 16 different unitary operations by Bob to complete the teleportation protocol. Alice communicates her measurement outcome ($\vert\Phi^{xy}\rangle_1 \vert\Phi^{x'y'}_{q}\rangle_3  $) classically to Bob.  Bob does the corresponding unitary transformation $\hat{U}$ (given in the third column of Table~\ref{tb.1}) on state of photon \textbf{2} to get 
\begin{equation}
\vert\Psi\rangle_2 = \left(a\vert H\rangle_2 +b \vert V\rangle_2\right)(\alpha \vert E\rangle_2 + \beta \vert O \rangle_2).
\end{equation}
which completes the teleportation of unknown state given in Eq.~\eqref{8}.

A SWAP operation is equivalent to consecutive $^pC_o$, $^oC_p$ and $^pC_o$ gate operations. Since the operation done in polarization will be transferred to the OAM after the SWAP operation, one can perform all the operations in polarization which are well known. Two half wave plates before and after the SWAP gate can perform 16 unitary operations which is required for the teleportation. A circuit diagram of the proposed scheme is given in Fig.~\ref{fg.5}.
\begin{table*}[!ht]
\begin{center}
  \begin{tabular}{| c | l | l | }
    \hline
    Measurement Outcome  & State of Bob's Photon  & Unitary Transformations ($\hat{U}$) \\ \hline
   $\vert\phi^+\phi^+\rangle$& $\left(\alpha \vert V\rangle +\beta \vert H\rangle \right)\left(a\vert O\rangle +b\vert E\rangle \right)$ &$ \left[ I^o \otimes \sigma^p_1\right] * U_{SWAP}* \left[\sigma^p_1 \otimes I^o\right] $  \\ \hline
   $\vert\phi^+\phi^-\rangle$& $\left(\alpha \vert V\rangle -\beta \vert H\rangle \right)\left(a\vert O\rangle +b\vert E\rangle \right)$ &$ \left[ I^o \otimes \sigma^p_1\right] * U_{SWAP}* \left[i\sigma^p_2 \otimes I^o\right] $  \\ \hline
   $\vert\phi^-\phi^+\rangle$& $\left(\alpha \vert V\rangle +\beta \vert H\rangle \right)\left(a\vert O\rangle - b\vert E\rangle \right)$ &$ \left[ I^o \otimes i\sigma^p_2\right] * U_{SWAP}* \left[\sigma^p_1 \otimes I^o\right] $  \\ \hline
   $\vert\phi^-\phi^-\rangle$& $\left(\alpha \vert V\rangle -\beta \vert H\rangle \right)\left(a\vert O\rangle -b\vert E\rangle \right)$ &$ \left[ I^o \otimes i\sigma^p_2\right] * U_{SWAP}* \left[i\sigma^p_2 \otimes I^o\right] $ \\ \hline
   $\vert\psi^+\psi^+\rangle$& $\left(\alpha \vert H\rangle +\beta \vert V\rangle \right)\left(a\vert E\rangle +b\vert O\rangle \right)$ &$ \left[ I^o \otimes I^p\right] * U_{SWAP}* \left[I^p \otimes I^o\right] $ \\ \hline
   $\vert\psi^+\psi^-\rangle$& $\left(-\alpha \vert H\rangle +\beta \vert V\rangle \right)\left(a\vert E\rangle +b\vert O\rangle \right)$ &$ \left[ I^o \otimes I^p\right] * U_{SWAP}* \left[\sigma^p_3 \otimes I^o\right] $ \\ \hline
   $\vert\psi^-\psi^+\rangle$& $\left(\alpha \vert H\rangle +\beta \vert V\rangle \right)\left(a\vert E\rangle -b\vert O\rangle \right)$ &$ \left[ I^o \otimes \sigma^p_3\right] * U_{SWAP}* \left[I^p \otimes I^o\right] $ \\ \hline
   $\vert\psi^-\psi^-\rangle$& $\left(-\alpha \vert H\rangle +\beta \vert V\rangle \right)\left(a\vert E\rangle -b\vert O\rangle \right)$ &$ \left[ I^o \otimes \sigma^p_3\right] * U_{SWAP}* \left[\sigma^p_3 \otimes I^o\right] $ \\ \hline
   $\vert\phi^+\psi^+\rangle$& $\left(\alpha \vert H\rangle +\beta \vert V\rangle \right)\left(a\vert O\rangle +b\vert E\rangle \right)$ &$ \left[ I^o \otimes \sigma^p_1\right] * U_{SWAP}* \left[I^p \otimes I^o\right] $ \\ \hline
   $\vert\phi^+\psi^-\rangle$& $\left(-\alpha \vert H\rangle +\beta \vert V\rangle \right)\left(a\vert O\rangle +b\vert E\rangle \right)$ &$ \left[ I^o \otimes \sigma^p_1\right] * U_{SWAP}* \left[\sigma^p_3 \otimes I^o\right] $ \\ \hline
   $\vert\phi^-\psi^+\rangle$& $\left(\alpha \vert H\rangle +\beta \vert V\rangle \right)\left(a\vert O\rangle -b\vert E\rangle \right)$ &$ \left[ I^o \otimes i\sigma^p_2\right] * U_{SWAP}* \left[I^p \otimes I^o\right] $ \\ \hline
   $\vert\phi^-\psi^-\rangle$& $\left(-\alpha \vert H\rangle +\beta \vert V\rangle \right)\left(a\vert O\rangle -b\vert E\rangle \right)$ &$ \left[ I^o \otimes i\sigma^p_2\right] * U_{SWAP}* \left[\sigma^p_3 \otimes I^o\right] $ \\ \hline
   $\vert\psi^+\phi^+\rangle$& $\left(\alpha \vert V\rangle +\beta \vert H\rangle \right)\left(a\vert E\rangle +b\vert O\rangle \right)$ &$ \left[ I^o \otimes I^p\right] * U_{SWAP}* \left[\sigma^p_1 \otimes I^o\right] $ \\ \hline
   $\vert\psi^+\phi^-\rangle$& $\left(\alpha \vert V\rangle -\beta \vert H\rangle \right)\left(a\vert E\rangle +b\vert O\rangle \right)$ &$ \left[ I^o \otimes I^p\right] * U_{SWAP}* \left[i\sigma^p_2 \otimes I^o\right] $ \\ \hline
   $\vert\psi^-\phi^+\rangle$& $\left(\alpha \vert V\rangle +\beta \vert H\rangle \right)\left(a\vert E\rangle -b\vert O\rangle \right)$ &$ \left[ I^o \otimes I^p\right] * U_{SWAP}* \left[i\sigma^p_2 \otimes I^o\right] $ \\ \hline
   $\vert\psi^-\phi^-\rangle$& $\left(\alpha \vert V\rangle -\beta \vert H\rangle \right)\left(a\vert E\rangle -b\vert O\rangle \right)$ &$ \left[ I^o \otimes \sigma_3^p\right] * U_{SWAP}* \left[i\sigma^p_2 \otimes I^o\right] $ \\ \hline
  \end{tabular}
\caption {Wave function corresponding to Bob's photon and the required unitary transformation corresponding to Alice's 
measurement outcome.$\sigma^p_1, \sigma^p_2, \sigma^p_3$ are Pauli matrices for polarization, $I^p$ and $I^o$ are identity matrices for polarization and OAM. }\label{tb.1}
\end{center}
\end{table*}

\subsection{Experimental realization of $^pC_o$, $^oC_p$ gates and SOBA}
\label{exp}
\begin{figure}[h]
  \begin{center}
    \includegraphics[width=2in]{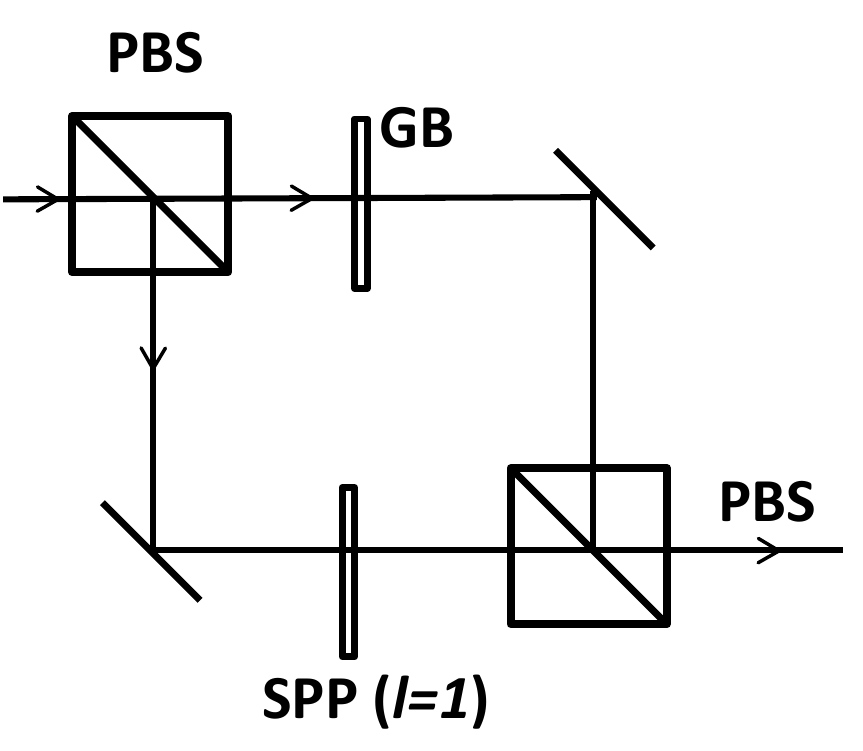}
    \caption{Experimental scheme for the implementation of polarization controlled OAM  $C_{NOT}$ gate ($^pC_o$). PBS - polarizing beam splitter, SPP - spiral phase plate, GB - glass block.}\label{fg.b} 
  \end{center}  
\end{figure}

A Polarization controlled OAM $C_{NOT}$ gate can be implemented using a modified Mach-Zehnder interferometer where the normal beam splitter is replaced by a polarizing beam splitter (PBS) as shown in Fig.~\ref{fg.b}. The reflected arm of the interferometer, through which the vertically polarized photons travel, contains a spiral phase plate (SPP) of order 1 which will convert the even OAM state to odd and vice versa. On the other arm with horizontally polarized photons, the  OAM state remains unchanged. These states superpose at the second PBS and emerges as a single $^pC_o$ gate output. A glass block is introduced to compensate the extra phase introduced by the SPP. 

\begin{figure}[h]
  \begin{center}
    \includegraphics[width=3in]{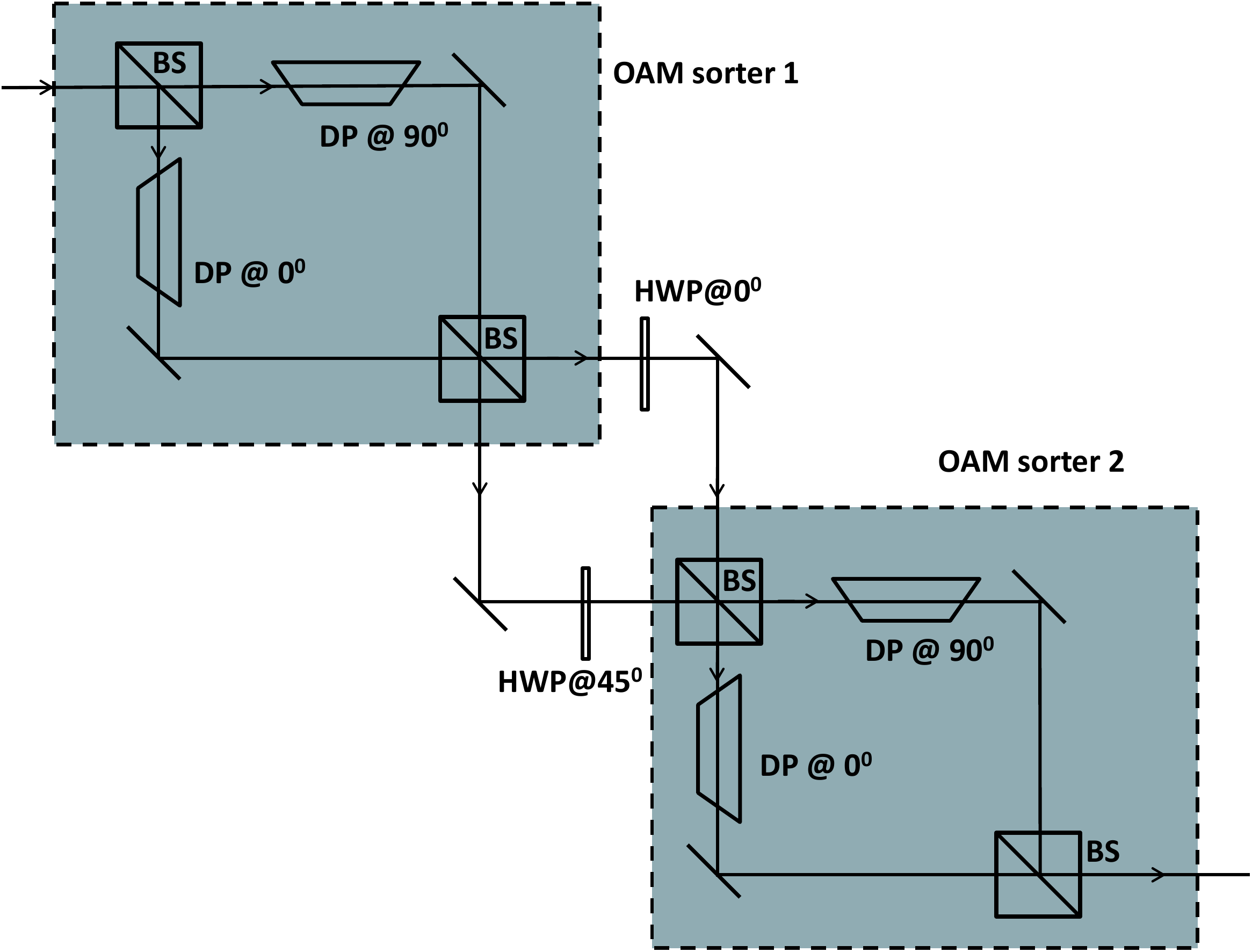}
    \caption{(Color online) Experimental scheme for the implementation of OAM controlled polarization $C_{NOT}$ gate ($^oC_p$). DP - dove prism, BS - 50:50 beam splitter, HWP - half wave plate}\label{fg.a} 
  \end{center}
\end{figure}

 We pursue a similar method to construct the OAM controlled polarization $C_{NOT}$ gate. For that, we need to use a beam splitter which transmits the photons with even OAM state and reflects the photons with odd OAM state. This can be achieved by an OAM sorter~\cite{Leach}. We replace the two PBSs in the previous setup with OAM sorters and instead of the SPP we use a half wave plate at $45^0$ in one arm that transforms the polarization state to its orthogonal state. The schematic of the setup is given in Fig.~\ref{fg.a}. The OAM sorter is another Mach-Zehnder interferometer containing dove prisms in each arm with a relative angle of $90^0$.    

\begin{figure}[h]
  \begin{center}
    \includegraphics[width=3in]{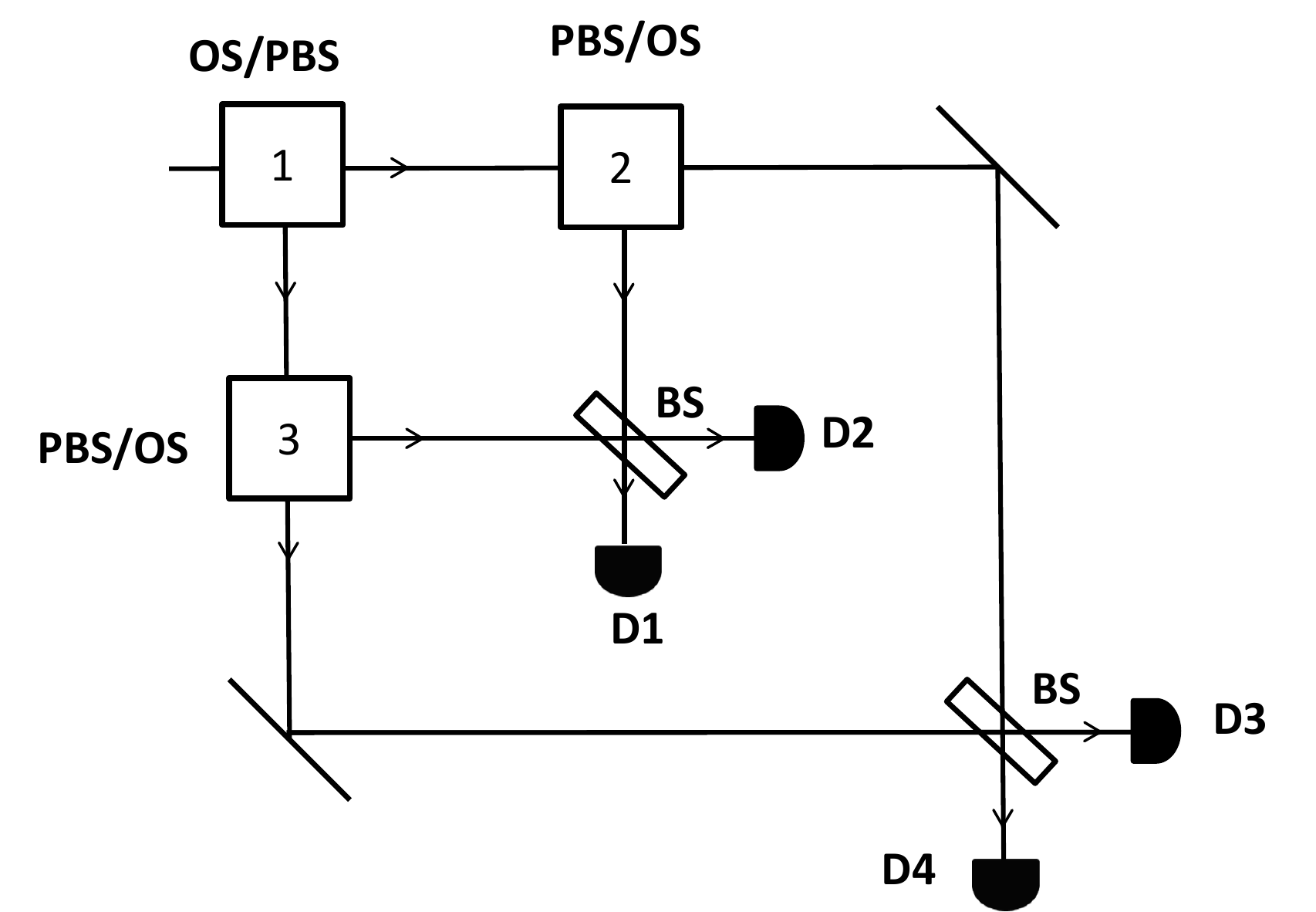}
    \caption{Experimental scheme for the implementation of SOBA in photon \textbf{1/3}. OS - OAM sorter, PBS - polarizing beam splitter, BS - 50:50 beam splitter}\label{fg.c} 
  \end{center}
\end{figure}

 The spin-orbit Bell state analysis with OAM in the even/odd states is explained by Khoury and Milman \cite{Khoury2011}. However, in the present protocol, the SOBA in photon \textbf{1} and photon \textbf{3} differ slightly in their implementation. As shown in the Fig.~\ref{fg.c}, two Mach-Zehnder interferometers with OAM sorters (OS) and polarizing beam splitters with photon detectors can implement SOBA. For  the photon \textbf{1}, the block 1 the Fig.~\ref{fg.c} is an OAM sorter and the blocks 2 and 3 are polarizing beam splitters. For SOBA in photon \textbf{3}, the block 1 is a PBS while the blocks 2 and 3 are OAM sorters. Detection on each of the detector will indicate the corresponding spin-orbit Bell state of the  photon. 

Here are the advantages of the described teleportation scheme 
\begin{itemize}

\item[(i)] As evident from the scheme, the number of particles required to teleport two independent qubits are reduced by 25\% by taking advantage of an extra DOF per photon.

\item[(ii)] We use single particle two-qubit Bell measurement instead of two particle joint Bell measurements. Its implementation is experimentally simple and one can achieve 100\% efficient Bell sate measurement and hence the teleportation.

\item[(iii)] In our scheme Bob need not to do unitary transformations on the OAM, since the same can be implemented by operations on the polarization before and after the SWAP.

\item[(iv)] The described three particle hyper-entangled states can be utilized for a multi-party teleportation scheme with two senders and a common receiver. Here Alice, and Charlie, with no entanglement channel between them, share photons \textbf{1} and \textbf{3}. The combined polarization-OAM quantum state of Alice's and Charlie's photons will be teleported to Bob who carries photon \textbf{2}.  
\end{itemize}

\section{Quantum Key Distribution}\label{sc.5}
Theoretically, the entanglement based QKD protocols are 
equivalent to the non-entanglement based ones (such as BB84). However, in 
practice, due to the strong quantum correlation and 
non-locality~\cite{jennewein2,bouwmeester1,ursin,gisin2,naik}, the entanglement 
based protocols are regarded more useful (for example, they have intrinsic 
randomness of the distributed key and extremely low probability of double 
photons). In recent years, entanglement in two degrees of freedom such as 
polarization and OAM has led to interesting applications in 
cryptography~\cite{mafu,complete}. 

 Here we are proposing a QKD protocol using the entangled system described in section~\ref{sc.2}. We take a similar state as in Eq.~\eqref{2} given by
\begin{equation}
\label{5}\nonumber \vert \Psi\rangle_{123} = \vert \xi_p\rangle_1 \otimes(\vert 00\rangle_{12}+\vert 1-1 \rangle_{12}+\vert -11 \rangle_{12}) \otimes \\ (\vert HH\rangle_{23}+\vert VV\rangle_{23} )\vert\chi_{o}\rangle_3,
\end{equation}	
where $ \vert 0\rangle$, $ \vert 1\rangle$ and $ \vert -1\rangle$ are the OAM states of photons.
Here we take a three dimensional subspace of infinite dimensional OAM entangled state produced by SPDC process with Gaussian beam as pump instead of even/odd entangled state.

\begin{figure}[h]
  \begin{center}
    \includegraphics[width=3in]{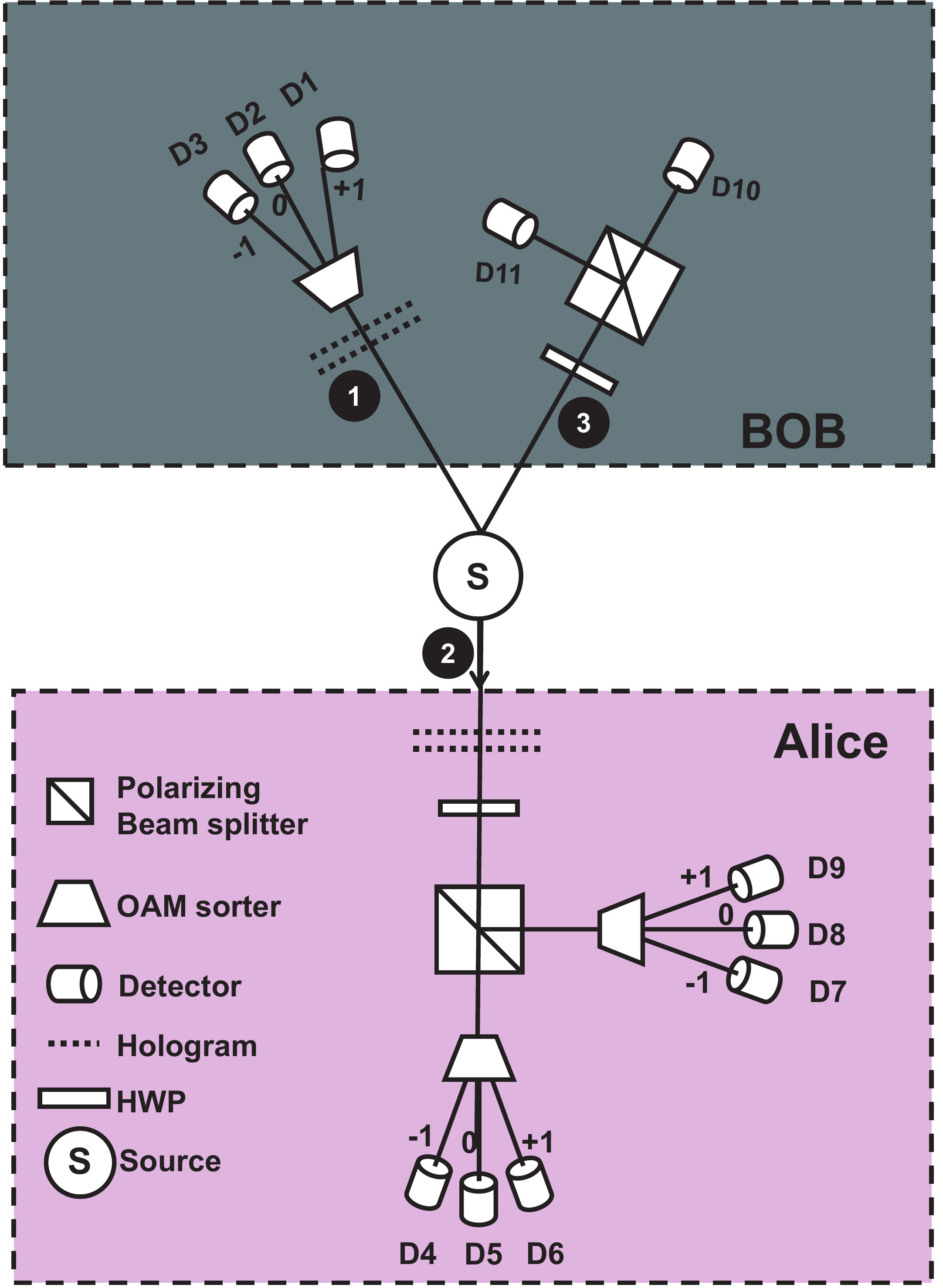}
    \caption{(Color online) Schematic diagram for quantum key distribution between Alice and Bob. Photons 1-2 have OAM entanglement and 1-3 have polarization entanglement. $D_{i}$ - detectors, HWP - half wave plate, PBS - polarizing beam splitter. \label{fg.6}}
  \end{center}
\end{figure}  

The experimental implementation of the protocol is shown in Fig.~\ref{fg.6}. 
Alice will receive photon \textbf{2} and Bob will receive photons \textbf{1} and \textbf{3} respectively. Photon \textbf{2} is entangled with photon \textbf{3} in polarization and it is entangled with photon \textbf{1} in OAM. The photons \textbf{1} and \textbf{3} do not have any correlation. 
Alice will measure both OAM and Polarization states, Bob will measure polarization of photon \textbf{3} and OAM of photon \textbf{1}.
 
Alice and Bob randomly measure polarization states of their respective photons in the following set of basis ($\gamma_i  $ = 0$^\circ$, 22.5$^\circ$, 45$^\circ$, 67.5$^\circ$),  ($ \delta_i$ = 22.5$^\circ$, 45$^\circ$, 67.5$^\circ$, 180$^\circ$) respectively. These measurements can be done by a half wave plate and polarizing beam splitter. Note that $\gamma_i/2$ and $\delta_i/2 $ are the fast axis orientation angles of the half wave plates. Each angle represents a basis which is used for measurement and corresponding to each of them, there are two measurement outcomes. The pairs of angles used by Alice and Bob for which the sum is  0$^\circ$ and  180$^\circ$  will give perfect correlation between them. The data corresponding to these correlated photons have same bits and  can be used for the key. Alice and Bob will compare their polarization measurement basis for the key distribution as well as for the security of the key. After sufficient number of measurements 4/16 of the data are useful for key, two sets of 4/16 of the data are used to check CHSH inequalities (S and S$^\prime$) and the remaining 4/16 are discarded due to unmatched bases.

For the OAM correlation of entangled photons, we follow the key sharing scheme  used in~\cite{groblacher}. In this scheme, Alice and Bob measure the OAM state of their photons in three different randomly chosen bases $A_{1}$, $A_{2}$, $A_{3}$ and $B_{1}$, $B_{2}$, $B_{3}$ respectively.  This is done by a pair of shifted holograms and an OAM sorter. The set of bases ($A_{3}$ and $B_{3}$) has perfect correlation and the coincidence corresponding to it can be used to generate the key. Alice and Bob will compare their hologram settings for QKD and security check. In total, there are 9 possible measurements. After taking sufficient number of measurements, 1/9 of the produced data can be used for the key. For Bell-type inequality check, 4/9 of the data will be used which confirms the security of the key and the remaining 4/9 of the data are redundant.

The security of the protocol is mainly checked by the violation of Bell's inequality test. All the three parties should check Bell like inequality with their data in order to check eavesdropping. If there is a violation of inequality, entanglement is preserved and there is no eavesdropping in the channel. The CHSH  parameters S and S$^\prime$ for photons entangled in the polarization DOF are given by \cite{chsh}
\begin{equation}
 S=  E(\gamma_{1},\delta_{1})-E(\gamma_{1},\delta_{3})+E(\gamma_{3},\delta_{1})+E(\gamma_{3},\delta_{3}) 
\end{equation}
\begin{equation}
 S^\prime=  E(\gamma_{2},\delta_{2})+E(\gamma_{2},\delta_{4})+E(\gamma_{4},\delta_{2})-E(\gamma_{4},\delta_{4})
\end{equation}
with
\begin{equation}
\hspace{-.2cm} E(\gamma_i ,\delta_j) = \frac{R_{12}(\gamma_i,\delta_j)+R_{1^\prime2^\prime}(\gamma_i,\delta_j)-R_{12^\prime}(\gamma_i,\delta_j)-R_{1^\prime2}(\gamma_i,\delta_j)}{R_{12}(\gamma_i,\delta_j)+R_{1^\prime2^\prime}(\gamma_i,\delta_j)+R_{12^\prime}(\gamma_i,\delta_j)+R_{1^\prime2}(\gamma_i,\delta_j)} \hspace{-.1cm}
\end{equation}
where $R_{12},R_{1^\prime 2^\prime},R_{12^\prime}$ and $R_{1^\prime 2}$ are the coincidences $P(D11,D9+D8+D7),\ P(D10, D4+D5+D6),\ P(D11, D4+D5+D6)$ and $P(D10,D9+D8+D7)$ respectively.

For any local realistic theory, the CHSH parameters $S, S^\prime \le 2$. Non-local nature of the entanglement will violate any of these inequalities and this violation can be used for checking the security of the key shared by Alice and Bob. The combination of $\gamma$ and $\delta$ for Bell's inequality test and for the key distribution is given in Table \ref{tb.2}.

\begin{table}
\begin{center}
  \begin{tabular}{| c  l  l l l| }
  \hline
Alice              & $\gamma_{1}$ & $\gamma_{2}$ & $\gamma_{3}$ & $\gamma_{4}$ \\
\hline
Bob\hspace{-0.03cm} \vline  & & & & 
\\
$\delta_{1}$ \ \  \hspace{0.0001cm} \vline     & S & Key & S & $\times$ \\
$\delta_{2}$ \ \ \ \vline    & $\times$ & S$^\prime$ & Key & S$^\prime$ \\
$\delta_{3}$ \ \ \ \vline    & S &$\times$ & S & Key  \\
$\delta_{4}$ \ \ \ \vline    & Key & S$^\prime$ & $\times$ & S$^\prime$  \\
\hline
\end{tabular}
\caption{Description of data usage corresponding to Alice's and Bob's measurement angles. Here S and S$^\prime$ are used for security check through CHSH inequality and $\times$ is the discarded data.} \label{tb.2}
\end{center}
\end{table}

Photons \textbf{1} and \textbf{2}  have OAM correlation, so it will violate the following Bell's inequality for 3 dimensional case \cite{groblacher,collins}
 \begin{eqnarray}
 \label{7} S =  P(A_{1}= B_{1})+P(A_{2}=B_{1}-1)+ P(A_{2}=B_{2})+\nonumber\\P(A_{1}=B_{2}) 
-P(A_{1}=B_{1}-1)-P(A_{2}=B_{1})+\nonumber\\P(A_{2}=B_{2}-1)+ P(A_{1}=B_{2}+1) 
 \le  2, 
  \end{eqnarray}
  where
 \begin{equation}
  \label{9} P(A_{a}=B_{b}+k)=\sum_{j=0}^{2}P(A_{a}=j,B_{b}=(j+k) Mod\ 3).
 \end{equation}
 
The shifts of holograms are chosen in such a way that they maximally violate the Bell-type inequality. $j$=0,1,2 corresponds to the detection of OAM states 0, 1 and -1 respectively. The coincidence measurements with the combinations of $P(D1, D6+D9)$, $P(D1, D5+D8)$, $P(D2, D4+D7)$, $P(D2, D6+D9)$, $P(D2, D5+D8)$, $P(D2, D4+D7)$, $P(D3, D6+D9)$, $P(D3, D5+D8)$ and $P(D3, D4+D7)$ are required for the key distribution and  to check the security of the key using Eq.~\eqref{7} and~\eqref{9}.

Our protocol has three advantages compared to traditional 
Ekert protocol~\cite{ekert1991}.
\begin{itemize}
\item[(i)] If Ekert protocol is used, then around $4n$ pairs 
of photons, i.e., $8n$ photons (in practice, little more than $8n$) are 
required to establish a secret key of length $n$ between 
Alice and Bob. In our approach, around $6n$ photons would be required for the 
same purpose. Hence, in terms of number of photons required, Ekert's
protocol is 33\% less efficient than ours.

\item[(ii)] On the other hand, if the same number of photons are used in Ekert 
and our protocol and if the target key length is also the same, then because
of more entanglement resource, our protocol would have more redundancy and hence
can tolerate more noise. It is easy to see that the security in each degree
of freedom is equivalent to that of the Ekert protocol.

\item[(iii)] Further, this state can be used for multi-party QKD. Photons 1, 2 
and 3 can be distributed amongst Alice, Bob and Charlie respectively.
Now, two sets of independent keys can be generated, one for Alice-Bob and 
another for Alice-Charlie.
\end{itemize}

\section{Conclusion}\label{sc.6}
Possibility of a new entangled state and its application in teleporting a two-qubit OAM-polarization quantum state and in the QKD have been discussed. The new teleportation method overcomes the difficulty of measuring in two particle polarization Bell basis, by implementing independent single particle two-qubit Bell measurements. The method critically depends on the experimental realizations of polarization $C_{NOT}$ gate as well as single particle two-qubit $C_{NOT}$  gates. All the 16 unitary transformations which are required for this teleportation scheme can be realized with linear optical components along with a SWAP gate. Once the state given in Eq.~\eqref{4} is achieved, one can have a 100\% efficient teleportation of two simultaneous qubits. Extending this to higher dimensions is mathematically trivial though creating entanglement in different DOFs has experimental limitations. With the new QKD protocol, the sender and the receiver need to use less resource than traditional Ekert protocol to share the secret key of the same length. Multi-party schemes also can be developed for the teleportation and the QKD using the described protocols. 

\section{Acknowledgments}
Authors wish to acknowledge Prof. Anirban Pathak, JIIT Noida for fruitful discussions and suggestions.

\label{}




\begin{thebibliography}{10}
\newcommand{\enquote}[1]{``#1''}

\bibitem{bennett1993}
C.~H. Bennett, G.~Brassard, C.~Cr\'epeau, R.~Jozsa, A.~Peres, and W.~K.
  Wootters, {Teleporting an unknown quantum state via dual classical and
  einstein-podolsky-rosen channels,} Phys. Rev. Lett. {70}, 1895--1899
  (1993).

\bibitem{bennett1992}
C.~H. Bennett and S.~J. Wiesner, {Communication via one- and two-particle
  operators on einstein-podolsky-rosen states,} Phys. Rev. Lett. {69},
  2881--2884 (1992).

\bibitem{ekert1991}
A.~K. Ekert, {Quantum cryptography based on bell's theorem,} Phys. Rev. Lett.
  {67}, 661--663 (1991).

\bibitem{Bouwmeester1997}
D.~Bouwmeester, J.-W. Pan, K.~Mattle, M.~Eibl, H.~Weinfurter, and A.~Zeilinger,
  {Experimental quantum teleportation,} Nature {390}, 575--579 (1997).

\bibitem{riebe}
M.~Riebe, H.~H{\"a}ffner, C.~Roos, W.~H{\"a}nsel, J.~Benhelm, G.~Lancaster,
  T.~K{\"o}rber, C.~Becher, F.~Schmidt-Kaler, D.~James \emph{et~al.},
  {Deterministic quantum teleportation with atoms,} Nature {429},
  734--737 (2004).

\bibitem{kim}
Y.-H. Kim, S.~P. Kulik, and Y.~Shih, {Quantum teleportation of a polarization
  state with a complete bell state measurement,} Phys. Rev. Lett. {86},
  1370--1373 (2001).

\bibitem{wei}
T.-C. Wei, J.~T. Barreiro, and P.~G. Kwiat, {Hyperentangled bell-state
  analysis,} Phys. Rev. A {75}, 060305 (2007).

\bibitem{barreiro2008}
J.~T. Barreiro, T.-C. Wei, and P.~G. Kwiat, {Beating the channel capacity limit
  for linear photonic superdense coding,} Nature Phys. {4}, 282--286
  (2008).
  
  \bibitem{Sheng1}
Y.-B. Sheng, F.-G. Deng, and G.~L. Long, {Complete hyperentangled-bell-state
  analysis for quantum communication,} Phys. Rev. A {82}, 032318 (2010).

\bibitem{Aolita}
L.~Aolita and S.~P. Walborn, {Quantum communication without alignment using
  multiple-qubit single-photon states,} Phys. Rev. Lett. {98}, 100501
  (2007).

\bibitem{Souza}
C.~E.~R. Souza, C.~V.~S. Borges, A.~Z. Khoury, J.~A.~O. Huguenin, L.~Aolita,
  and S.~P. Walborn, {Quantum key distribution without a shared reference
  frame,} Phys. Rev. A {77}, 032345 (2008).

\bibitem{Santos}
B.~C. dos Santos, K.~Dechoum, and A.~Z. Khoury, {Continuous-variable
  hyperentanglement in a parametric oscillator with orbital angular momentum,}
  Phys. Rev. Lett. {103}, 230503 (2009).

\bibitem{Borges}
C.~V.~S. Borges, M.~Hor-Meyll, J.~A.~O. Huguenin, and A.~Z. Khoury, {Bell-like
  inequality for the spin-orbit separability of a laser beam,} Phys. Rev. A
  {82}, 033833 (2010).

\bibitem{Lixiang}
L.~Chen and W.~She, {Single-photon spin-orbit entanglement violating a
  bell-like inequality,} J. Opt. Soc. Am. B {27}, A7--A10 (2010).

\bibitem{Khoury2011}
A.~Z. Khoury and P.~Milman, {Quantum teleportation in the spin-orbit variables
  of photon pairs,} Phys. Rev. A {83}, 060301 (2011).

\bibitem{Walborn2004}
S.~P. Walborn, A.~N. de~Oliveira, R.~S. Thebaldi, and C.~H. Monken,
  {Entanglement and conservation of orbital angular momentum in spontaneous
  parametric down-conversion,} Phys. Rev. A {69}, 023811 (2004).

\bibitem{Leach}
J.~Leach, M.~J. Padgett, S.~M. Barnett, S.~Franke-Arnold and J.~Courtial, {Measuring the Orbital Angular Momentum of a Single Photon,} Phys. Rev. Lett. {88}, 25901 (2002).

\bibitem{simon}
R.~Simon and N.~Mukunda, {Minimal three-component su (2) gadget for
  polarization optics,} Phys. Lett. A {143}, 165--169 (1990).

\bibitem{Reddy}
S.~G. Reddy, S.~Prabhakar, A.~Aadhi, A.~Kumar, M.~Shah, R.~Singh, and R.~Simon,
  {Measuring the mueller matrix of an arbitrary optical element with a
  universal su (2) polarization gadget,} J. Opt. Soc. Am. A {31}, 610--615 (2014).

\bibitem{pati}
A.~K. Pati and P.~Agrawal, {Probabilistic teleportation of a qudit,} Phys. Lett. A {371}, 185--189 (2007).

\bibitem{goyal}
S.~K. Goyal, P.~E. Boukama-Dzoussi, S.~Ghosh, F.~S. Roux, and T.~Konrad,
  {Qudit-teleportation for photons with linear optics,} Sci. Rep.
  {4}, 4543 (2014).

\bibitem{Goyal2013}
S.~K. Goyal and T.~Konrad, {Teleporting photonic qudits using multimode quantum
  scissors,} Sci. Rep. {3}, 3548 (2013).

\bibitem{babichev2003}
S.~A. Babichev, J.~Ries, and A.~I. Lvovsky, {Quantum scissors: Teleportation of
  single-mode optical states by means of a nonlocal single photon,} Europhys. Lett. {64}, 1 (2003).

\bibitem{jennewein2}
T.~Jennewein, C.~Simon, G.~Weihs, H.~Weinfurter, and A.~Zeilinger, {Quantum
  cryptography with entangled photons,} Phys. Rev. Lett. {84},
  4729--4732 (2000).

\bibitem{bouwmeester1}
D.~Bouwmeester and A.~Zeilinger, {The physics of quantum information: basic
  concepts,} in \enquote{The physics of quantum information,}  (Springer,
  2000), pp. 1--14.

\bibitem{ursin}
R.~Ursin, F.~Tiefenbacher, T.~Schmitt-Manderbach, H.~Weier, T.~Scheidl,
  M.~Lindenthal, B.~Blauensteiner, T.~Jennewein, J.~Perdigues, P.~Trojek
  \emph{et~al.}, {Entanglement-based quantum communication over 144 km,} Nature
  Phys. {3}, 481--486 (2007).

\bibitem{gisin2}
N.~Gisin, G.~Ribordy, W.~Tittel, and H.~Zbinden, {Quantum cryptography,} Rev.
  Mod. Phys. {74}, 145--195 (2002).

\bibitem{naik}
D.~S. Naik, C.~G. Peterson, A.~G. White, A.~J. Berglund, and P.~G. Kwiat,
  {Entangled state quantum cryptography: Eavesdropping on the ekert protocol,}
  Phys. Rev. Lett. {84}, 4733--4736 (2000).

\bibitem{mafu}
M.~Mafu, A.~Dudley, S.~Goyal, D.~Giovannini, M.~McLaren, M.~J. Padgett,
  T.~Konrad, F.~Petruccione, N.~L\"utkenhaus, and A.~Forbes,
  {Higher-dimensional orbital-angular-momentum-based quantum key distribution
  with mutually unbiased bases,} Phys. Rev. A {88}, 032305 (2013).

\bibitem{complete}
V.~D'Ambrosio, E.~Nagali, S.~P. Walborn, L.~Aolita, S.~Slussarenko,
  L.~Marrucci, and F.~Sciarrino, {Complete experimental toolbox for
  alignment-free quantum communication,} Nat. Commun. {3}, 961
  (2012).

\bibitem{groblacher}
S.~Gröblacher, T.~Jennewein, A.~Vaziri, G.~Weihs, and A.~Zeilinger,
  {Experimental quantum cryptography with qutrits,} New J. Phys.
  {8}, 75 (2006).

\bibitem{chsh}
J.~F. Clauser, M.~A. Horne, A.~Shimony, and R.~A. Holt, {Proposed experiment to
  test local hidden-variable theories,} Phys. Rev. Lett. {23}, 880--884
  (1969).

\bibitem{collins}
D.~Collins, N.~Gisin, N.~Linden, S.~Massar, and S.~Popescu, {Bell inequalities
  for arbitrarily high-dimensional systems,} Phys. Rev. Lett. {88},
  040404 (2002).

\end{thebibliography}


\end{document}